\pdfoutput=1

\documentclass[aps,prd,reprint,groupedaddress, longbibliography]{revtex4-2}

\usepackage[left]{lineno}
\usepackage{longtable}
\usepackage{graphicx}
\usepackage{amssymb}
\usepackage{fancyvrb}
\usepackage{verbatimbox}
\usepackage{listings}
\usepackage{array}
\usepackage{longtable}
\usepackage{amsmath}
\usepackage{amssymb}
\usepackage{enumitem}
\usepackage[title]{appendix}
\usepackage{nicefrac,xfrac}

\begin{document}


\newcolumntype{P}[1]{>{\centering\arraybackslash}p{#1}}

\title{Study of the ionization efficiency for nuclear recoils in pure crystals}

\author{Y. Sarkis, Alexis Aguilar-Arevalo and Juan Carlos D'Olivo}

\affiliation{Instituto de Ciencias Nucleares, Universidad Nacional Aut\'onoma de M\'exico,  04510 CDMX, M\'exico}

\date{\today}

\begin{abstract}

We study the basic integral equation in Lindhard's theory describing the energy given to atomic motion by nuclear recoils in a pure material when the atomic binding energy is taken into account.
The numerical solution, which depends only on the slope of the velocity-proportional electronic stopping power and the binding energy, leads to an estimation of the ionization efficiency which is in good agreement with the available experimental measurements for Si and Ge. In this model, the {\it quenching factor} for nuclear recoils features a cut-off at an energy equal to twice the assumed binding energy. We argue that the model is a reasonable approximation for Ge  even for energies close to the cut-off, while for Si is valid up to recoil energies greater than $\sim500$~eV. 

\end{abstract}

\keywords{quenching factor, dark matter, nuclear recoils}

\maketitle

\section{Introduction}

In experiments dedicated to the detection of rare events producing low energy depositions ($\sim10$~keV or less), {\it e.g.}, direct dark matter (DM) searches or the detection of coherent neutrino-nucleus scattering (CEN$\nu$S), the experimental signal generally entails the detection of the recoiling target ions following a scattering event. The amount of electronic excitation produced by a recoiling ion is typically smaller than that produced by a recoiling electron of the same energy, 
this has been commonly referred to as {\it quenching}.  The usage of this term may not be in general well justified, since no loss of the elementary electronic excitations ({\it total quanta}) is required to occur in order to explain this difference in all cases. Nonetheless, for simplicity and consistency with current literature, hereafter in this work, we will use the term {\it quenching factor}  (QF) to refer to the ionization efficiency for pure crystals, like Si and Ge.


In 1963, Lindhard {\it et al.} \cite{lindhard:1963} developed a theoretical model that has been used to explain this quenching, aimed at describing energy depositions of the order of a few keV or higher, when atomic binding energies can be safely neglected. After more than 50 years, the original formulation by Lindhard and collaborators (hereafter referred to as Lindhard's, in short) remains widely in use, and has shown to be successful at describing measurements in this energy regime.
As experiments have lowered their detection thresholds reliably observing energy depositions well below 1~keV, understanding the QF at those low energies has become crucial to estimate their sensitivities to the physical models they aim to test.

Recent measurements of the QF for nuclear recoils in silicon (Si) \cite{chavarria:2016, izraelevitch:2017} exhibit a clear deviation from the Lindhard model for energies below 4~keV, while data for germanium (Ge) \cite{Ge1,Ge2,Ge3,Ge4,Ge5,Ge6} are in good agreement. 

In a recent article, Sorensen \cite{Sorensen} aimed to obtain a QF valid at lower energies by bringing back the atomic binding energy into Lindhard's original simplified equation. He estimated this binding to be of the order of the electron-hole pair creation energy ($\sim 3$~eV for Si and Ge), and his solutions exhibit a cut-off of the order of one to a few hundred eV. This result is troublesome \cite{Gascon}, since 
it is not obvious how a low binding energy could produce such a high threshold in the QF. The present work was partially motivated by this observation, and will show that, when properly incorporated into the model, a constant binding energy results in a cut-off in the QF at a value of the same order of magnitude.

This paper is organized as follows. 
In Section \ref{sec:lindhard-model} we give a brief summary of the ideas in Lindhard's theory arriving at the simplified integral equation describing the energy given to ions by a recoiling ion in a homogeneous medium, and his equation for the QF, when the binding energy is neglected. 
In Section \ref{sec:simp-integ-eq} we discuss the changes that are needed in order to maintain the binding energy in the model to the lowest order, arriving at a modified version of the simplified integral equation. We propose a simple
ansatz 
for the solution depending on two new parameters, besides the electronic stopping constant $k$ already introduced by Lindhard. We end this section with a description of the numerical solution which depends only on $k$ and the binding energy, 
which works well in the low energy regime for Ge and Si, most relevant for current and future low-threshold DM ({\it e.g.},  \cite{COGENT,CDMSlite,CDMS,DAMIC,DAMICM})  and CE$\nu$NS ({\it e.g.}, \cite{CONNIE,conus,TEXexp}) experiments.
In Section \ref{sec:fits-to-data} we fit the QF obtained from both, the approximate and numerical solutions, to experimental measurements for Si and Ge  to find the relevant parameters in each case. 
The conclusions are presented in Section \ref{sec:conclusions}.

\section{The Lindhard model}
\label{sec:lindhard-model}

When an ion in a homogeneous substance moves with a kinetic energy $E$, heading towards the collision with another ion in the material, after recoiling off an interaction with an incident particle ({\it e.g.} the coherent scattering of a neutrino or a DM particle), is sets off a cascade of slowing-down processes that dissipate this energy throughout the medium. If the ion recoils from the interaction with the incident particle with an energy $E_R$ and the energy $U$ is lost to disruption of atomic bonding,
then $E_R=E+U$. Note that, under the assumption of an elastic collision, $E_R$ corresponds to the
kinetic energy transferred by the incident particle to the struck ion in the material.
A sudden approximation is made for the collision, where the timescale of elastic collision is much smaller than timescale of atomic processes.
Lindhard's theory \cite{lindhard:1963} concerns with determining the fraction of $E_R$ which is given to electrons, $H$, and that which is given to translational motion of ions, $N$, assuming $E_R = H+N$. This separation can be written in terms of 
reduced dimensionless quantities as 
\begin{equation}
    \varepsilon_R=\eta+\nu ,
\end{equation}
\noindent 
where $\varepsilon_R=c_{\scriptscriptstyle Z}E_R$, $\eta=c_{\scriptscriptstyle Z}H$, and $\nu=c_{\scriptscriptstyle Z}N$, and the scaling factor $c_{\scriptscriptstyle Z}=11.5/Z^{7/3}$~keV$^{-1}$ is defined for a medium with a single atomic species of atomic number $Z$.

The model is simplified by considering the equations obeyed by the average quantities $\bar{\eta}$ and $\bar{\nu}$, for which appropriate probability distributions are assumed to exist, and such that $\varepsilon_R = \bar\eta+\bar\nu$.
%

%
It is reasonable to assume that $\bar\eta$ represents an upper limit to the available signal in a detector operating in ionization-only mode, such as those used for extreme low-mass WIMP searches and CE$\nu$NS detection.
The {\it nuclear quenching factor} is defined as the fraction of the total energy deposited by the incident particle which is transferred to the electrons
\begin{equation}
\label{eq:quenching-factor-1}
    f_n = \frac{\bar\eta}{\varepsilon_R}
        = \frac{\varepsilon + u - \bar\nu}{\varepsilon + u} ,
\end{equation}

\noindent
where $u=c_{\scriptscriptstyle Z}U$.

\begin{figure}[t]
\centering
\scalebox{0.4}{\includegraphics{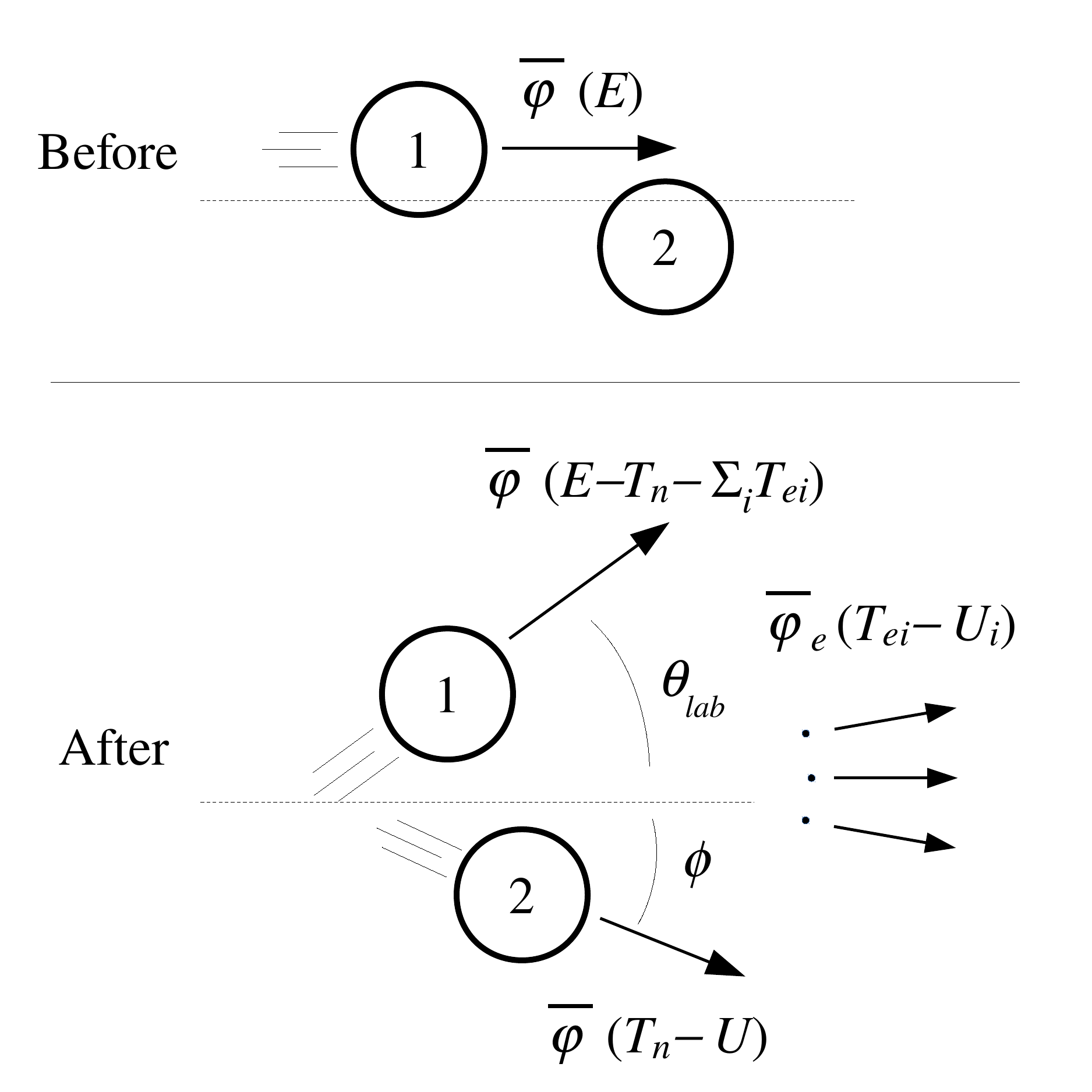}}
\caption{\small Scattering of a recoiling ion in the lab frame. The average physical effect of the recoiling ion $\bar\varphi(E)$ equals the sum of the average physical effects of the struck ion, the ejected electrons, and itself, after the collision. $U_i$ is the ionization energy to free electron $i$. The other quantities are described in the text.}
\label{fig:sketch}
\end{figure}

Lindhard considered any physical quantity $\varphi$ (of which $\eta$ and $\nu$ are examples) that is additive over the individual slowing-down processes spawned by the initial scattering.
Suppose that a recoiling ion, with kinetic energy $E$, strikes an ion in the medium transferring the energy $T_n$ to its center of mass, and the energy $T_{ei}$ to each ionized electron. If $U$, in Lindhard's own words, is the energy spent in {\it ``disrupting the atomic binding"}, then the additivity of $\varphi$ is encoded in the basic integral equation
\begin{eqnarray}\label{Eq:BasIntEq}
\int d\sigma_{n,e}\left[  \bar\varphi(E-T_{n}-\Sigma_i T_{ei}) +
\bar\varphi(T_{n}-U) -\bar\varphi(E) \right. \nonumber \\ 
\left.
+\Sigma_i\bar\varphi_e(T_{ei}-U_{ei})
\right] =0,
\end{eqnarray}

\noindent
where $\sigma_{n,e} $ is the effective cross-section for the interaction of the recoiling ion with the ions or electrons in the medium and integration over $\int d\sigma_{n,e}$ represents the sum over all possible interactions (impact parameters).
In the last term, $\bar\varphi_e$ is the function describing the contribution of ejected electrons to $\bar\varphi$, each with ionization energy $U_{ei}$.
Eq.(\ref{Eq:BasIntEq}) states that the average physical effect caused by the initial recoiling ion before the collision, $\bar\varphi(E)$, equals the sum of the average physical effects caused by the ion, the struck ion, and the ejected electrons after the collision. This situation is depicted in Fig.~\ref{fig:sketch}.
In general, $U$ is not limited to the energy needed to remove the ion from its site, but it can also include contributions to excitation or ionization of bound atomic electrons, and therefore incorporates the Migdal effect \cite{migdal1,migdal2} into the model. In scintillating materials, electronic excitation can be a very significant component of the total signal. In principle, an additional term for the contribution of excited electronic states could be added to Eq.(\ref{Eq:BasIntEq}), but we will not consider it in the treatment presented here.

Lindhard used five basic approximations in order to cast Eq.(\ref{Eq:BasIntEq}), for $\bar\varphi(E)=\bar\nu(E)$, in a simplified form for which he found an approximate numerical solution, expected to be valid for sufficiently large energies:
(A) ionized electrons do not produce atomic recoils with appreciable energy, hence the term $\sum_i\bar\varphi_{e}(T_{ei}-U_{ei})$ can be dropped; (B)  neglect the atomic binding $U$ under the assumption that it is in general smaller than the energy transferred to the recoiling ions, hence $\varepsilon_R\approx\varepsilon$; (C) the energy transferred to ionized electrons is also small compared to that transferred to recoiling ions; (D) the effects of electronic and atomic collisions can be treated separately; (E) $T_n$ is also small compared to the energy $E$.

The interactions between recoiling ions are modeled as two-body elastic scatterings of identical particles in a screened Coulomb potential $V(r)=(e^2Z^2/r)\phi_0(r/a)$. Here, $\phi_0(r/a)$ is the single atom Thomas-Fermi screening function \cite{Thomasfermi} with a corrected length scale $a=0.8853\,a_0/(Z^{1/3}\sqrt{2})$, and $a_0$ is the Bohr radius. With this model Lindhard found that the atomic scattering cross section could be written as $d\sigma_n = dt f(t^{1/2})/2t^{3/2}$, where $t=\varepsilon^2\sin^2(\theta/2)$, $\theta$ is the scattering angle in the center of mass, and $f(t^{1/2})$ is a function only of $t$.

The electronic stopping power can be expressed as $1/N_e(dE/dR)_e=\int d\sigma_e (\Sigma_i T_{ei})$ \cite{wilson:1977}, where $N_e$ is the electron number density and $R$ is the distance travelled by an ionizing projectile. It appears naturally as a consequence of approximations (C) and (D), and in terms of the reduced quantities $\varepsilon$ and $\rho = \pi a^2 N_e R$, can be written as 
\begin{equation}
\label{eq:elect_stopping}
    S_e(\varepsilon) = d\varepsilon/d\rho = k\varepsilon^{1/2},
\end{equation}

\noindent
where $k=0.133\,Z^{2/3}A^{-1/2}$. This velocity proportionality of the electronic stopping power appears to hold in a variety of substances, from gaseous to semiconductor targets, although indications of a threshold velocity 
below which a projectile loses no energy to electrons are known to exist \cite{Stp}.

Putting all these approximations together, including $u=0$ (approximation B), Lindhard arrived at his  simplified integral equation for the average energy given to atomic motion
\begin{eqnarray}
\label{Eq:SimpIntEq}
\nonumber
k\;\varepsilon^{1/2} \bar\nu'(\varepsilon) &=& \int^{\varepsilon^{2}}_{0}dt\frac{f(t^{1/2})}{2t^{3/2}} \\
&&\times\left[ \bar\nu(\varepsilon-t/\varepsilon) +\bar\nu(t/\varepsilon) -\bar\nu(\varepsilon) \right].
\end{eqnarray}

\noindent
He found an approximate numerical solution of Eq.(\ref{Eq:SimpIntEq}) imposing the boundary condition that $\bar\nu(\varepsilon)\rightarrow \varepsilon$ when $\varepsilon\rightarrow 0$ (and noting that $\bar\nu''(\varepsilon)<0$), from where the QF in Eq.(\ref{eq:quenching-factor-1}) can be calculated as
\begin{equation}\label{Eq:LindQF:def}
f_{n}=\frac{\bar\eta(\varepsilon)}{\varepsilon}=\frac{\varepsilon-\bar\nu(\varepsilon)}{\varepsilon},
\end{equation}

\noindent 
which he parametrized in the following way
\begin{eqnarray}
\nonumber
\bar\nu_{L}(\varepsilon)&=&\frac{\varepsilon}{1+kg(\varepsilon)}, \;\;\;\bar\nu(\varepsilon)\equiv\bar\nu_{L}(\varepsilon)\\
g(\varepsilon)&=& 3\varepsilon^{0.15} + 0.7\varepsilon^{0.6}  + \varepsilon. 
\label{Eq:LindSol:nu}
\end{eqnarray}

\noindent
The last expression is well known to the experimental community studying low energy depositions by nuclear recoils.

It is interesting to note that there is an inconsistency with the boundary condition imposed by Lindhard which, on one hand implies that $\bar\nu_L'(\varepsilon)\rightarrow 1$ when $\varepsilon \rightarrow 0$, as stated above, while on the other, by applying L'Hopital's rule directly to Eq.(\ref{Eq:SimpIntEq}) it can be shown that $\lim_{\varepsilon \rightarrow 0} \bar\nu_L'(\varepsilon)= 0$, hinting at the existence of a discontinuity in the first derivative at zero.
Despite its limitations, Lindhard's model has been very successful in describing the QF for nuclear recoils in Si up to $\varepsilon\gtrsim 0.1$ (4~keV), and so far all available data for Ge, corresponding to $\varepsilon\gtrsim$ 0.00088 (250~eV).

\section{Simplified integral equation with binding energy}
\label{sec:simp-integ-eq}

We wish to find a version of the simplified integral equation, Eq.(\ref{Eq:SimpIntEq}), where approximation (B) has been removed in a mathematically consistent way. In Ref. \cite{Sorensen} this approximation was relaxed by replacing the term $\bar\nu(t/\varepsilon)$ with $\bar\nu(t/\varepsilon-u)$. While this is certainly part of the required modifications, attention must be paid to the lower limit of integration on the right-hand side of Eq.(\ref{Eq:SimpIntEq}), which should be set to $\varepsilon u$, as is suggested by not allowing the argument of $\bar\nu(t/\varepsilon-u)$ to become negative. The same lower limit can be recovered by modelling the atomic scattering as the collision of semi-hard spheres, as is shown in Appendix~\ref{App:semi-hard}.

In addition to bringing back the binding energy, in going from Eq.(\ref{Eq:BasIntEq}) to Eq.(\ref{Eq:SimpIntEq}), the term $\bar\varphi(E-T_n-\Sigma_i T_{ei})$ has been expanded to first order in $\Sigma_i T_{ei}/(E-T_n)\ll 1$, but, it has also been assumed that $T_n/E$ is small to some extent (approximation E). In the interest of finding a solution valid for lower energies ({\it e.g.} $\varepsilon>0.01$ in Si) we will perform a similar expansion, but keeping a term of order $T_n (\Sigma_i T_{ei})$, namely
\begin{eqnarray}
    \bar\nu(E-T_n-\Sigma_i T_{ei}) 
    &\approx& \bar\nu(E-T_n) - \bar\nu'(E)(\Sigma_i T_{ei}) \nonumber\\ 
    &&  
    + \bar\nu''(E) T_n (\Sigma_i T_{ei}) ,
\label{Eq:phiExp}
\end{eqnarray}

\noindent
where terms of order $(\Sigma_i T_{ei})^2$ or higher, have been dropped. 
The additional term proportional to $\bar\nu''(E)$ will have an important effect when assessing the accuracy of our approximate solution, and will be key to the implementation of the numerical solution.
Substituting Eq.(\ref{Eq:phiExp}) into Eq.(\ref{Eq:BasIntEq}), and integrating over the nuclear and electronic cross sections, putting also in effect approximation (D), we arrive at our proposed form of the modified simplified integral equation
\begin{eqnarray}\label{Eq:ModSimIntEq}
k\;\varepsilon^{1/2}\bar\nu'(\varepsilon) -\tfrac{1}{2}k\;\varepsilon^{3/2}\bar\nu''(\varepsilon)  =  
\int^{\varepsilon^{2}}_{\varepsilon u} dt \frac{f(t^{1/2})}{2t^{3/2}}
\hspace{1.5cm}
 \nonumber \\
\times\left[\bar\nu(\varepsilon-t/\varepsilon) + \bar\nu(t/\varepsilon -u)-\bar\nu(\varepsilon) \right], \hspace{0.4cm}
\end{eqnarray}

\noindent
where we have considered a mean value of the energy transferred to the struck ion $\bar{t}_n\approx \langle t_n\rangle = \langle \varepsilon \sin^2\theta/2 \rangle = {\textstyle \frac{1}{2}}\varepsilon$, (where $t_n=c_{\scriptscriptstyle Z}T_n=t/\varepsilon$) in order to recover the electronic stopping power from the integration of the second order term (see Appendix \ref{App:Deriv:ModSimIntEq}). 

In what follows we will use the “average” Molière-like screening function given in \cite{wilson:1977} for the determination of  $f(t^{1/2})$. Other screening functions are available \cite{Ziegler1985,J.Lensen}, and the differences between them can affect the determination of the constant $k$, but are expected to be constrained to the interval  0.1-0.2.

The model depicted in Fig.~\ref{fig:sketch} 
requires that prior to producing any effect the struck ion must recoil with a kinetic energy larger than $U$, otherwise the argument in $\bar\varphi(T_n-U)$ becomes negative.
Modeling the process as the collision of semi-hard spheres (see Appendix \ref{App:semi-hard}), we recognize $U$ as the depth of the {\it soft} part of the potential, and can be associated with the energy given to the electrons occupying the shells 
above the inner noble-like non-valence shells of the ion.
If sufficient energy is available the collision can induce excitation of  electrons from these shells, as well as from the valence to the conduction band, producing a number of electron-hole ({\it e--h}) pairs, and possibly also create a vacancy and self-intersticial (Frenkel) pair \cite{FrSi,FrGe} in the lattice.
In general $U$ will depend on the kinetic energy of the recoiling ion $E$.

Table \ref{tab:uvals} shows the values of the binding energies, relative to the top of the valence band, for electrons occupying inner shells above the $[{\rm Ne}]^2$ or $[{\rm Ar}]^{18}$ cores in Si and Ge, respectively \cite{atomicbinding1,atomicbinding2}. The table also lists the average {\it e--h} production energy and the dislocation energy (average energy to create a FrenkelFig.~ pair) for each element \cite{FrSi,FrGe}.
In Si, a recoiling ion (labeled 1 in Fig.~\ref{fig:sketch}) moving through the lattice with, say  $\varepsilon/c_{\scriptscriptstyle Z}=350$~eV of kinetic energy, could strike an ion (labeled 2 in Fig.~\ref{fig:sketch}) and cause an electron from its $2p$ shell to reach the conduction band (100 eV + a fraction of 3.7 eV), in addition to causing a handful more to reach it from the valence band. Depending on the number of excited electrons and their energies, the struck ion could also become dislocated from the lattice.
Similarly, in Ge, an ion moving with $\varepsilon/c_{\scriptscriptstyle Z}=50$~eV of kinetic energy could strike an ion and excite an electron from its $3d$ shell plus a few more from the valence band to the conduction band, or dislocate the ion. 
%
Note that the ion that initiates the cascade will also have lost, to atomic processes, some of the recoil energy with which it emerged from the interaction with the incoming particle  $\varepsilon_{\scriptscriptstyle R}$ (Migdal effect \cite{migdal1,migdal2}).

In the remainder of this work we will take $u(\varepsilon)=u$, a constant value, and explore its implications for the QF for nuclear recoils at low energies.

\begin{table}[t]
    \centering
    \caption{Binding energies, relative to the top of the valence band, for atomic shells between the noble core and the outer valence orbitals \cite{atomicbinding1,atomicbinding2}, average {\it e--h} creation energies, and dislocation energies \cite{FrSi,FrGe} in Si and Ge.}
    \begin{tabular*}{\columnwidth}{@{\extracolsep{\fill}} ccc|ccc} \hline
    \multicolumn{3}{c}{silicon} & \multicolumn{3}{c}{germanium}       \\ \hline
    Shell & $U$ (eV)  & $\# e$ &  Shell  & $U$ (eV)& $\# e$ \\ \hline
    $[{\rm Ne}]^4$  &  & 4 &  $[{\rm Ar}]^{18}$ &  & 18 \\
    $2p$            &  100 & 6 &   $3d$      &   30  & 10    \\  \hline
    Average $e$--$h$   & 3.7  & 4 & Average $e$-$h$ & 3.0 & 4      \\ 
    Dislocation     & 36   &   & Dislocation  & 23  &        \\ \hline
    \end{tabular*}
    \label{tab:uvals}
\end{table}

\subsection{Model with a constant $u$}
\label{sec:constant_u}

\noindent 
When $u$ is constant, Eq.(\ref{Eq:ModSimIntEq}) is only applicable for $\varepsilon \geq u$, otherwise the lower limit of integration derived from the semi-hard sphere model becomes ill-defined (see Appendix \ref{App:semi-hard}). 
Furthermore, since the right-hand side ({\it r.h.s.}) of Eq.(\ref{Eq:ModSimIntEq}) is the contribution to $\bar{\nu}$ from the recoiling ion (labeled 1 in Fig.~\ref{fig:sketch}), it must be non-negative for any $\varepsilon \geq u$.  Defining the quantity in square brackets in the integrand as
\begin{equation}
    I(\varepsilon,t) =
    \bar\nu(\varepsilon-t/\varepsilon) + \bar\nu(t/\varepsilon -u)-\bar\nu(\varepsilon) \;,
\end{equation}

\noindent
two observations are in order:
(1) at $\varepsilon=u$ the {\it r.h.s.} of Eq.(\ref{Eq:ModSimIntEq}) is equal to zero, since the uper and lower limits of integration are equal, therefore, $I(\varepsilon,t)\geq 0$ (must be nonnegative) for any $\varepsilon \geq u$, and
(2) evaluating the {\it r.h.s.} at any value of $\varepsilon > u$ requires knowledge of the function  $\bar\nu(\varepsilon)$ for values of $\varepsilon<u$.
Note that observation (1) further implies that $\bar\nu(\varepsilon)$ has the following linear form in the region below $u$
\begin{equation}
    \label{eq:nu_low_energy}
    \bar\nu(\varepsilon) = \varepsilon+u = \varepsilon_R,
    \;\;\;
    {\rm for} \;\;\;
    \varepsilon \leq  u .
\end{equation}

We now use Eq.(\ref{eq:quenching-factor-1}) to calculate the QF with $\bar{\nu}(\varepsilon)$ as the solution to the problem posed in Eq.(\ref{Eq:ModSimIntEq}). From the requirement in Eq.(\ref{eq:nu_low_energy}), it is clear that the QF will vanish for $\varepsilon \leq u$, or equivalently, for $\varepsilon_{\scriptscriptstyle R}\leq 2u$.
In the limit $u=0$ we recover Lindhard's model and QF. The constant $u$ model is one in which no energy will go into the motion of ions unless the initial ion recoils with an energy  $\varepsilon_{\scriptscriptstyle R} > 2u$.

From the values in Table \ref{tab:uvals} we can expect that this model will produce a cut-off in the QF for Si at recoil energies of the order of 200-400~eV, while for Ge it can be expected at energies of the order of 30-60~eV.

\subsection{Interpolation from low to high $\varepsilon$ behavior}

\begin{figure}[t]
\centering
\includegraphics[scale=0.40]{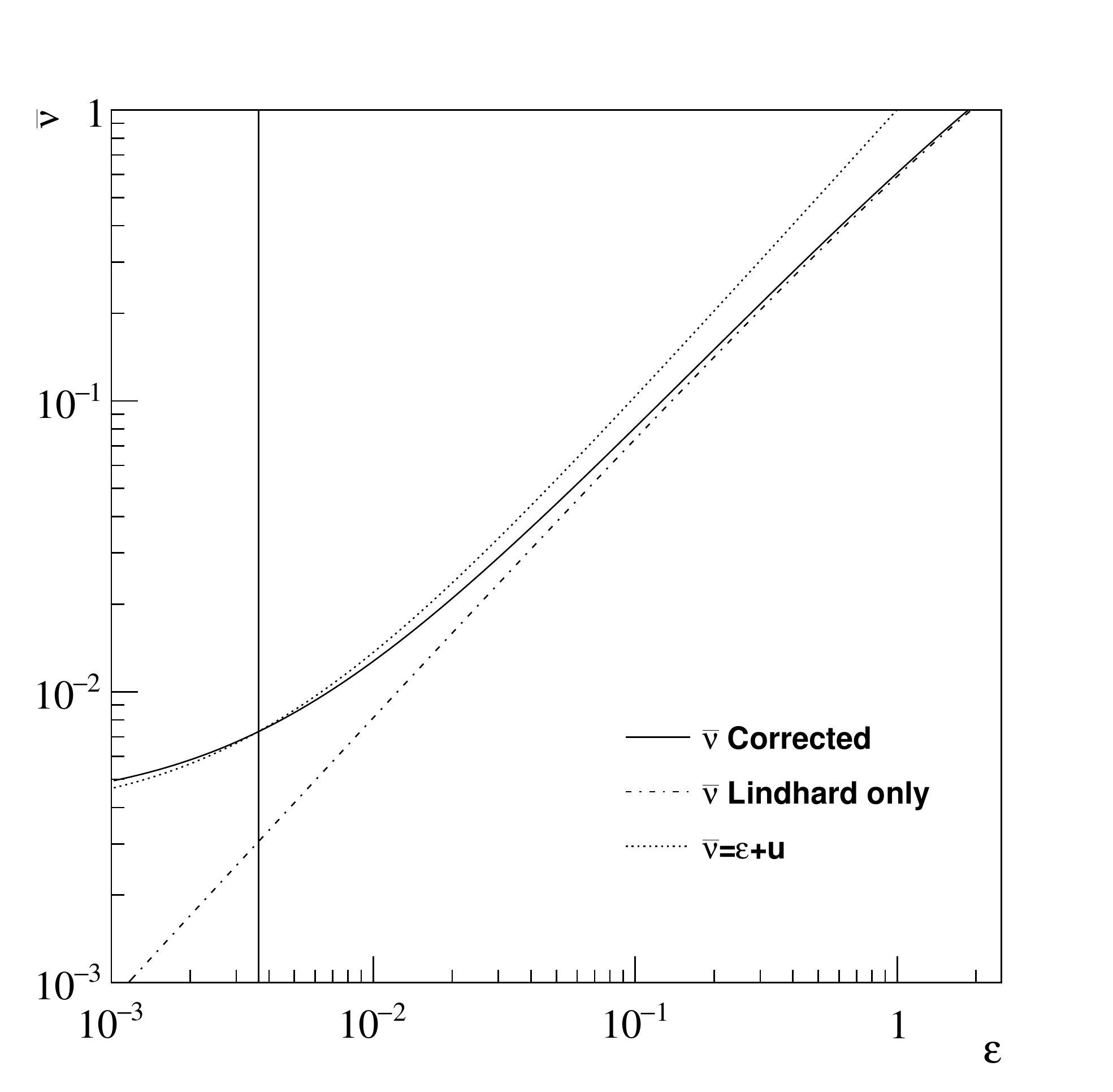}
\caption{ The function $\bar\nu(\varepsilon)$ from Eq.(\ref{eq:corrected_nu_v0}) fitted to the Si experimental data interpolates between the Lindhard solution at high energies, and the expected $\varepsilon+u$ (approximately) below $u$. A cut-off in the QF occurs at the crossing between $\bar\nu(\varepsilon)$ and $\varepsilon+u$ at $\varepsilon=u$ (vertical line).}
\label{fig:nu-comp-lind}
\end{figure}

It has been noted by some authors \cite{Gascon,Sorensen} that in Lindhard's original model the energy transferred to electrons is slightly overestimated. This is so because it primarily originates from the electronic stopping power of ions, assumed to be given by Eq.(\ref{eq:elect_stopping}), which vanishes at $\varepsilon=0$. However, if we consider that the effect of the binding energy is to suppress the energy transferred to electrons when the recoiling  ion has energies below $u$, we can argue that $\bar\eta$ needs to be corrected by a certain amount. If the correction is taken to be proportional to the electronic stopping power at energy $\varepsilon$ itself, plus a possible offset, we can write
\begin{equation}
    \bar\eta = \bar\eta_L - C_0\left( d\varepsilon/d\rho \right) - C_1,
\end{equation}
\noindent 
where $\bar\eta_L$ is the average energy transferred to electrons according to the Lindhard model. Since $\varepsilon = \bar\eta_L + \bar\nu_L$, the corrected average energy transferred to atomic motion is
\begin{equation}
\label{eq:corrected_nu_v0}
  \bar\nu = \bar\nu_L + C_0 \varepsilon^\frac{1}{2} + C_1 + u,
\end{equation}
\noindent
Notice that the model used in \cite{Sorensen} is equivalent to correcting 
$\bar\eta$ by a constant value, however, it is tested against Lindhard's  basic integral equation, Eq.(\ref{Eq:SimpIntEq}).
The general form in Eq.(\ref{eq:corrected_nu_v0}) can be made to approximately follow the required linear behavior expected near and below $u$, posited in Eq.(\ref{eq:nu_low_energy}), while at the same time coincide with Lindhard's solution at high $\varepsilon$, as can be seen in Fig.~\ref{fig:nu-comp-lind}. 
Such solution will produce a cut-off in the QF defined in Eq.(\ref{eq:quenching-factor-1}) at $\varepsilon = u$, provided that $\bar\nu(u)=2u$, and that $\bar\nu(\varepsilon)>\varepsilon + u$ for $\varepsilon<u$. 
One could also device a solution for $\bar\nu(\varepsilon)$ that equals $\varepsilon+u$ once $\varepsilon$ falls below $u$ by allowing it to have a discontinuity on the first derivative (a kink) at this value. 

As a way to measure the quality of our proposed solution we will follow \cite{Sorensen} and define the error

\begin{equation}
    \label{eq:error-metric}
    {\rm Error} = \left|
    \frac{  {r.h.s} -  {l.h.s} } { {r.h.s} +  {l.h.s} } 
    \right| \;,
\end{equation}

\noindent 
comparing the left-hand-side ($l.h.s$) and the right-hand-side ($r.h.s.$) of the modified integral equation, Eq.(\ref{Eq:ModSimIntEq}). As noted in \cite{Sorensen} evaluation of the $r.h.s.$ requires knowledge of the function $f(t^{1/2})$ to lower energies than considered by Lindhard. Therefore, we follow the useful prescription given therein and use the parametrization for the reduced nuclear stopping power $S_n(\varepsilon)$, Eq.(15) of \cite{wilson:1977}, to calculate $f(t^{1/2})$ by differentiation of $\varepsilon S_n(\varepsilon)$.

\subsection{Numerical solution}

From the observations in section \ref{sec:constant_u} we write the solution in the form:
\begin{equation}
    \bar{\nu}(\varepsilon) = 
    \left\lbrace
      \begin{array}{ll}
         \varepsilon + u       & ,\;\;\varepsilon < u \,\, ,\\
          \varepsilon + u -\lambda(\varepsilon)  & ,\;\;\varepsilon \geq u  \,\, ,
      \end{array}
    \right.
    \label{eq:splitsol}
\end{equation}

\noindent
where $\lambda(\varepsilon)$ is a continuous function satisfying $\lambda(u)=0$. In order for Eq.(\ref{eq:splitsol}) to be a solution to the integral equation, Eq.(\ref{Eq:ModSimIntEq}), $\lambda(\varepsilon)$ must have a discontinuity in its first (and therefore also in its second) derivative at $\varepsilon=u$. This is reminiscent of what happens in Lindhard's equation at $\varepsilon=0$, as mentioned at the end of Section \ref{sec:lindhard-model}. Defining these discontinuities as
\begin{eqnarray}
  \lim_{\zeta\rightarrow0} {\lambda'(u+\zeta)} = \alpha_1  , \;\;\;
  & \underset{\zeta \to 0}{\lim}{\lambda''(u+\zeta)} = \alpha_2,  \nonumber \\
  \lim_{\zeta\rightarrow0} \lambda'(u-\zeta) = 0 ,\;\;\;\;\; 
  & \underset{\zeta \to 0}{\lim} \lambda''(u-\zeta) = 0, \;\;
\end{eqnarray}
\noindent
with $\alpha_1\neq 0$ and $\alpha_2\neq 0$, consistently the condition to make the {\it l.h.s.} in Eq. (\ref{Eq:ModSimIntEq}) vanish at $\varepsilon=u$ is given by
\begin{equation}
    \alpha_1 = 1 + \tfrac{1}{2}\,u \alpha_2 .
    \label{eq:alpha1}
\end{equation}
\noindent
Therefore it is only necessary to determine one of the two parameters ({\it e.g.}, $\alpha_2$). In order for $\bar{\nu}(\varepsilon)$ to remain physical, its second and first derivatives  must satisfy the boundary conditions %
\begin{eqnarray}
    \lim_{\varepsilon\rightarrow \infty}
    \bar{\nu}''(\varepsilon) & = & 0^-  \;{\rm (from \; below)}, \;\;\;{\rm and}
   \label{eq:nuprime-limits-2}\\
    0\;\leq \bar{\nu}'(\varepsilon) &\leq& 1  \;\;\;\;
    {\rm for}\;\varepsilon\geq u , \label{eq:nuprime-limits-1} 
\end{eqnarray} 

\noindent
otherwise $\bar\nu$ will not match Lindhard's solution at high energies, if Eq.(\ref{eq:nuprime-limits-2}) is not satisfied, or the QF could become, either negative or greater than 1, if Eq.(\ref{eq:nuprime-limits-1}) is not satisfied.

For $\varepsilon=u$ the first condition, Eq.(\ref{eq:nuprime-limits-2}), restricts the possible values of $\alpha_2$ to lie in the interval 
\begin{equation}
    -2/u\leq \alpha_2 \leq 0 .
\label{eq:alpha2-limits}
\end{equation}

Given $u\neq 0$, and small values of the step size $h$, and tolerance $\delta$, (both $\ll 1$), we find a solution to Eq.(\ref{Eq:ModSimIntEq}) in the interval $u \leq\varepsilon\leq\varepsilon^{\rm max}$ by means of the following {\it shooting} method:
\begin{enumerate}

\item \label{step:start}
Set $\varepsilon^{\rm max}$ to a large initial value $\varepsilon^{\rm max}_0= 500~ u$, and the limits $\alpha_2^{\rm lo} = -2/u$, and $\alpha_2^{\rm hi} = 0$.

\item \label{step:sample-alpha2}
Sample a random value of $\alpha_2$ in the interval $\alpha_2^{\rm lo} \leq\alpha_2\leq\alpha_2^{\rm hi}$, calculate the corresponding value of $\alpha_1$ from Eq.(\ref{eq:alpha1}), and set the starting values
\begin{equation}
    \varepsilon_t=u, \;
    \lambda(u)=0,\; \lambda'(u)= \alpha_1,\; \lambda''(u)=\alpha_2 .
\end{equation}

\item \label{step:test-emax}
If $\varepsilon_t=\varepsilon^{\rm max}$, skip to step \ref{step:test-cond2}. Else, use Eq.(\ref{eq:splitsol}) to calculate $\bar{\nu}(\varepsilon_t)$, $\bar{\nu}'(\varepsilon_t)$, and $\bar{\nu}''(\varepsilon_t)$.

\item \label{step:test-cond1}
If the condition in Eq.(\ref{eq:nuprime-limits-1}) is satisfied, continue. Else, if it fails because $\bar{\nu}'(\varepsilon_t)<0$, set $\alpha_2^{\rm hi}=\alpha_2$, and return to \ref{step:sample-alpha2}. Else, if it fails because $\bar{\nu}'(\varepsilon_t)>1$, set $\alpha_2^{\rm lo}=\alpha_2$, and return to \ref{step:sample-alpha2}.

\item \label{step:calc-ggprime}
Calculate $\lambda(\varepsilon_t+h)$ and $\lambda'(\varepsilon_t+h)$ using a second order expansion of $\lambda$ about $\varepsilon_t$
\begin{eqnarray}
    \lambda(\varepsilon_t+h) & \approx & \lambda(\varepsilon_t) + \lambda'(\varepsilon_t) h \nonumber \\ 
        &     & \hspace{0.8cm} + \tfrac{1}{2}\lambda''(\varepsilon_t) h ^2 ,
        \nonumber \\
    \lambda'(\varepsilon_t+h) & \approx & \lambda'(\varepsilon_t) 
    + \lambda''(\varepsilon_t) h  ,
    \label{eq:g:secondorder}
\end{eqnarray}
\noindent
and calculate $\bar{\nu}(\varepsilon_t+h)$ and $\bar{\nu}'(\varepsilon_t+h)$.

\item \label{step:solveint}
Use Eq.(\ref{Eq:ModSimIntEq}) to solve for $\bar{\nu}''(\varepsilon_t+h)$, evaluating the integral in the {\it r.h.s.} numerically by interpolating the behavior of $\bar{\nu}(\varepsilon)$ between $u$ and $\varepsilon_t+h$ with cubic splines passing through all previous points.

\item \label{step:inc-et}
Set $\varepsilon_t$ to $\varepsilon_t+h$ and return to \ref{step:test-emax}

\item \label{step:test-cond2}
If the second derivative condition in Eq.(\ref{eq:nuprime-limits-2}) at $\varepsilon^{\rm max}$ is satisfied within a tolerance $\delta$, stop. Else, increment $\varepsilon^{\rm max}=\varepsilon^{\rm max}+\Delta$ and return to \ref{step:sample-alpha2}.

\end{enumerate}

\begin{figure}[b]
\centering
\includegraphics[scale=0.43]{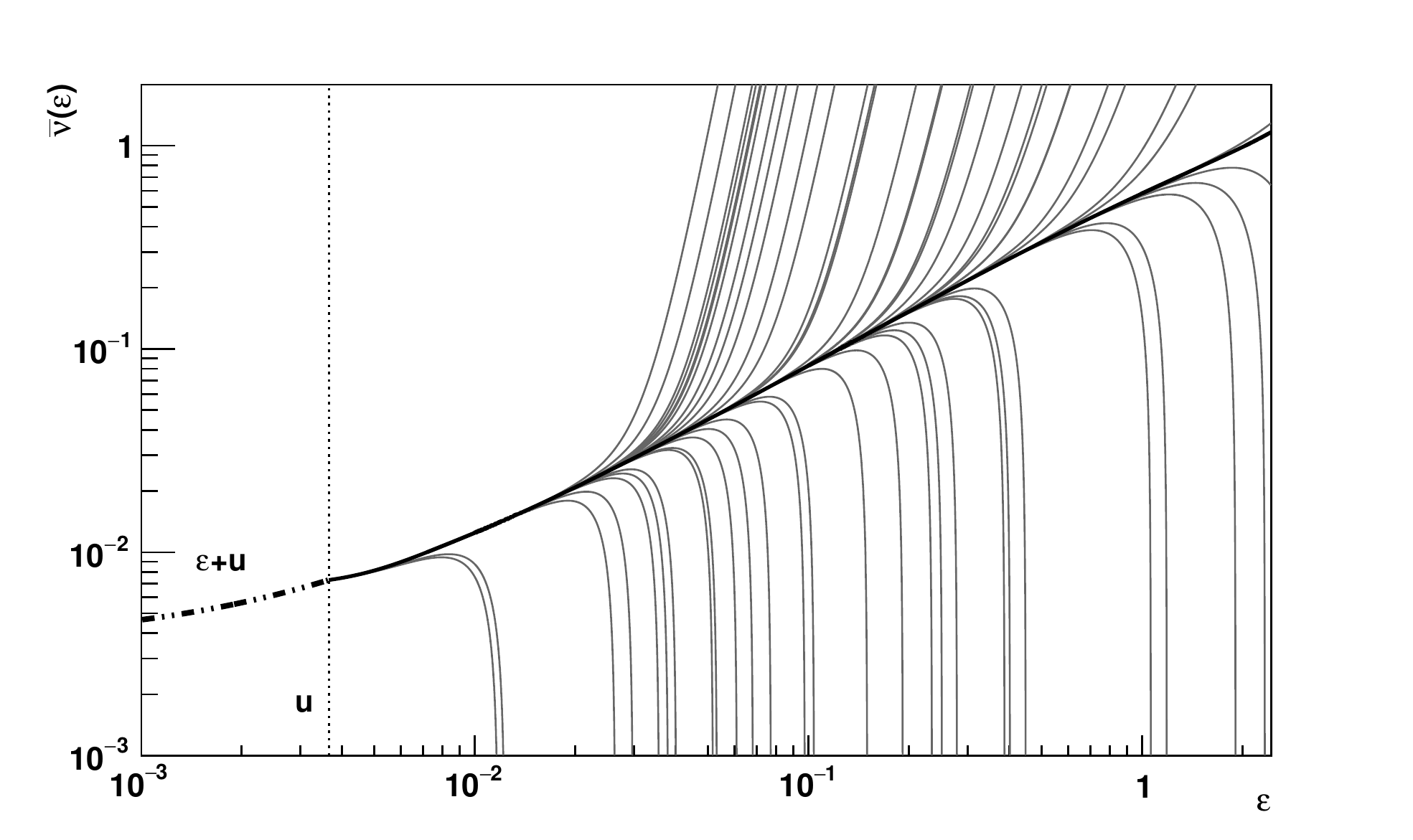}
\caption{Shooting method for Si. The red bold curve is the only one satisfying the boundary conditions in Eqs.(\ref{eq:nuprime-limits-1}) and (\ref{eq:nuprime-limits-2}).}
\label{fig:nu-shooting}
\end{figure}

 An example of the application of this method to the case of Si with $u=3.7\times10^{-3}$, and 1000 steps uniformly spaced in logarithmic scale in the interval $150$~eV$~<E_r<100~$~keV is illustrated in Fig.~\ref{fig:nu-shooting}. The second derivative condition in Eq.(\ref{eq:nuprime-limits-2}) is well satisfied at $\varepsilon$ corresponding to 100~keV, although for some values of $u$ and $k$, the condition  is satisfied at lower energies, for those cases Eq.(\ref{Eq:ModSimIntEq}) in step \ref{step:solveint} can be used without the second derivative term. The solutions from 61 random {\it shots} failing to satisfy the conditions in Eqs.(\ref{eq:nuprime-limits-1}) and (\ref{eq:nuprime-limits-2}) are shown as the black curves. The successful final shot satisfying the conditions in the interval of interest is shown in red.

\section{Fits to data }
\label{sec:fits-to-data}

\begin{table}[t]
    \centering
    \caption{Summary of the data sets used in this study.}
    \begin{tabular*}{\columnwidth}{@{\extracolsep{\fill}} ccc} \hline
         Data set & Energy range (keV) & \# points  \\ \hline
         \multicolumn{3}{c}{silicon}   \\ \hline
         Zech    \cite{Zech} &  4.30 - 53.7    &  8      \\
         Brian   \cite{Brian} &  4.15 - 75.7    &  4      \\
         CHICAGO \cite{chavarria:2016}     & 0.68 - 2.28  &  12  \\
         ANTONELLA \cite{izraelevitch:2017} & 1.79 - 20.67   & 14    \\ \hline
                  \multicolumn{3}{c}{germanium}    \\ \hline
         Jones (75) \cite{Ge4} & 0.254    &  1      \\
         COGENT     \cite{Ge7} & 0.65 - 1.22   &  4       \\
         TEXONO     \cite{Ge6} & 1.25 - 3.61   &  3       \\
         Messous    \cite{Ge9} & 2.71 - 8.72   &  3       \\ 
         Shutt      \cite{Ge10} & 17.50 - 70.05   &  7       \\ 
         Chassman   \cite{Ge2} & 10.04 - 73.17   &  16       \\     
                        
          \hline
    \end{tabular*}
    \label{tab:datasets}
\end{table}

The QF data sets used in this study are summarized in Table \ref{tab:datasets}. 
\begin{table}[b]
    \centering
    \caption{Fitted parameters for the ansatz in Eq.(\ref{eq:corrected_nu_v0}) for the different data sets. We report the binding energy $U=u/c_{\scriptscriptstyle Z}$. High $\chi^2/ndf$ reflect the tension among the data sets given the reported errors. The uncertainties are estimated so as to cover the variations among the data sets.}
    \begin{tabular*}{\columnwidth}{@{\extracolsep{\fill}} ccccc} \hline
           & $C_0$    & $C_1\,(\times10^{-5})$ & $U ($\rm keV$)$ & $\chi^2/ndf$ \\ \hline
        Si & ($9.1 \pm 4.4 )\times 10^{-3} $  & $3.33 \pm 1.2$ & $0.15 \pm 0.06 $  & 224/40 \\
        Ge & ($3.0 \pm 1.3)\times 10^{-4}$   & $0.62  \pm 0.12$  & $0.02 \pm 0.01$  & 56/35 \\
        \hline
    \end{tabular*}
    \label{tab:fitted-ansatz}
\end{table}
For Si, four data sets have been considered:  Zech \cite{Zech}, with 8 points in the energy range from 4.30 to 53.7 keV; Brian \cite{Brian}, with 4 points in the energy range from 4.15 to 75.7 keV; CHICAGO \cite{chavarria:2016} with 12 points in the energy range from 0.68 to 2.28 keV; ANTONELLA \cite{izraelevitch:2017} with 14 points in the energy range from 1.79 to 20.67 keV. The last two are the lowest energy measurements available to date.
For Ge, six data sets have been considered:  Jones (75) \cite{Ge4}, with 1 point at 0.254 keV; COGENT \cite{Ge7} with 4 points in the energy range from 0.65 to 1.22 keV; TEXONO \cite{Ge6} with 3 points in the energy range from 1.25 to 3.61 keV; Messous \cite{Ge9} with 3 points in the energy range from 2.71 to 8.72 keV; Shutt \cite{Ge10} with 7 points in the energy range from 17.50 to 70.05 keV; Chassman \cite{Ge2} with 16 points in the energy range from 10.04 to 73.17 keV.
%

\begin{table}[t]
    \centering
    \caption{Fitted parameters for the numerical solution to the different data sets. We report the binding energy $U=u/c_{\scriptscriptstyle Z}$. High $\chi^2/ndf$ reflect the tension among the data sets given the reported errors. The uncertainties are estimated so as to cover the variations among the data sets.}
   
    \begin{tabular*}{\columnwidth}{@{\extracolsep{\fill}} cccc} \hline
           & $k$    & $U ($\rm keV$)$ & $\chi^2/ndf$ \\ \hline
        Si & $0.161 \substack{+0.029 \\ -0.020}$  & $0.15 \substack{+0.10 \\ -0.05} $  & 349.2/40 \\ 
        Ge & $0.162 \substack{+0.028 \\ -0.021} $  & $0.02 \substack{+0.015 \\ -0.010} $  & 52.3/35 \\
        \hline
    \end{tabular*}
    \label{tab:fitted-numeric}
        \end{table}

\begin{figure}[b]
\includegraphics[scale=0.43]{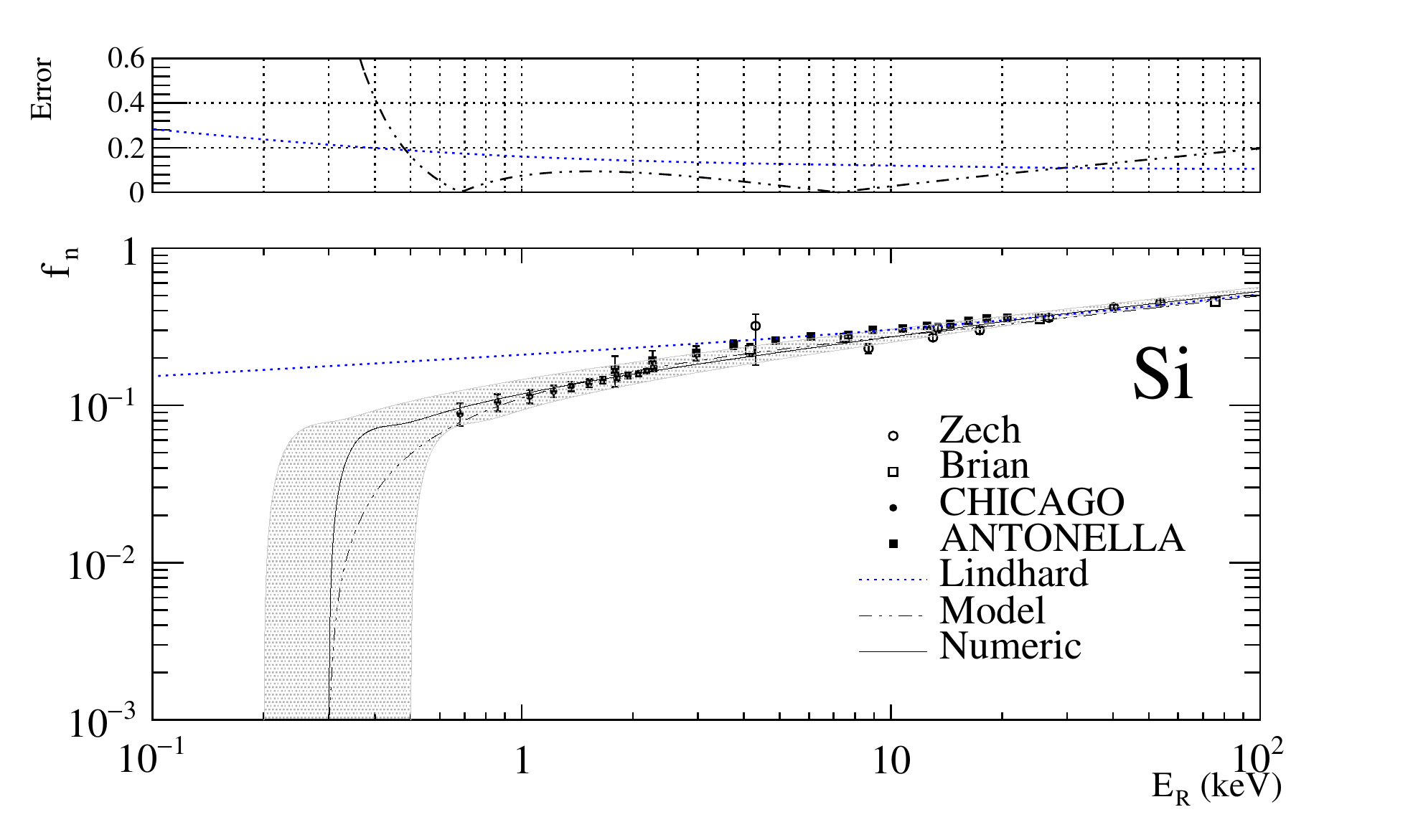}
\caption{ 
(Lower panel) Measurements of the QF in Si (points with error bars) compared tvo the Lindhard model (dot-dashed line), the ansatz of Eq.(\ref{eq:corrected_nu_v0}), and the numerical solution with $U=0.15$ keV and $k=0.161$. (Upper panel) Error in the ansatz and the Lindhard original model.
}
\label{fig:fit-Si}
\end{figure}

\begin{figure}[t]
\includegraphics[scale=0.41]{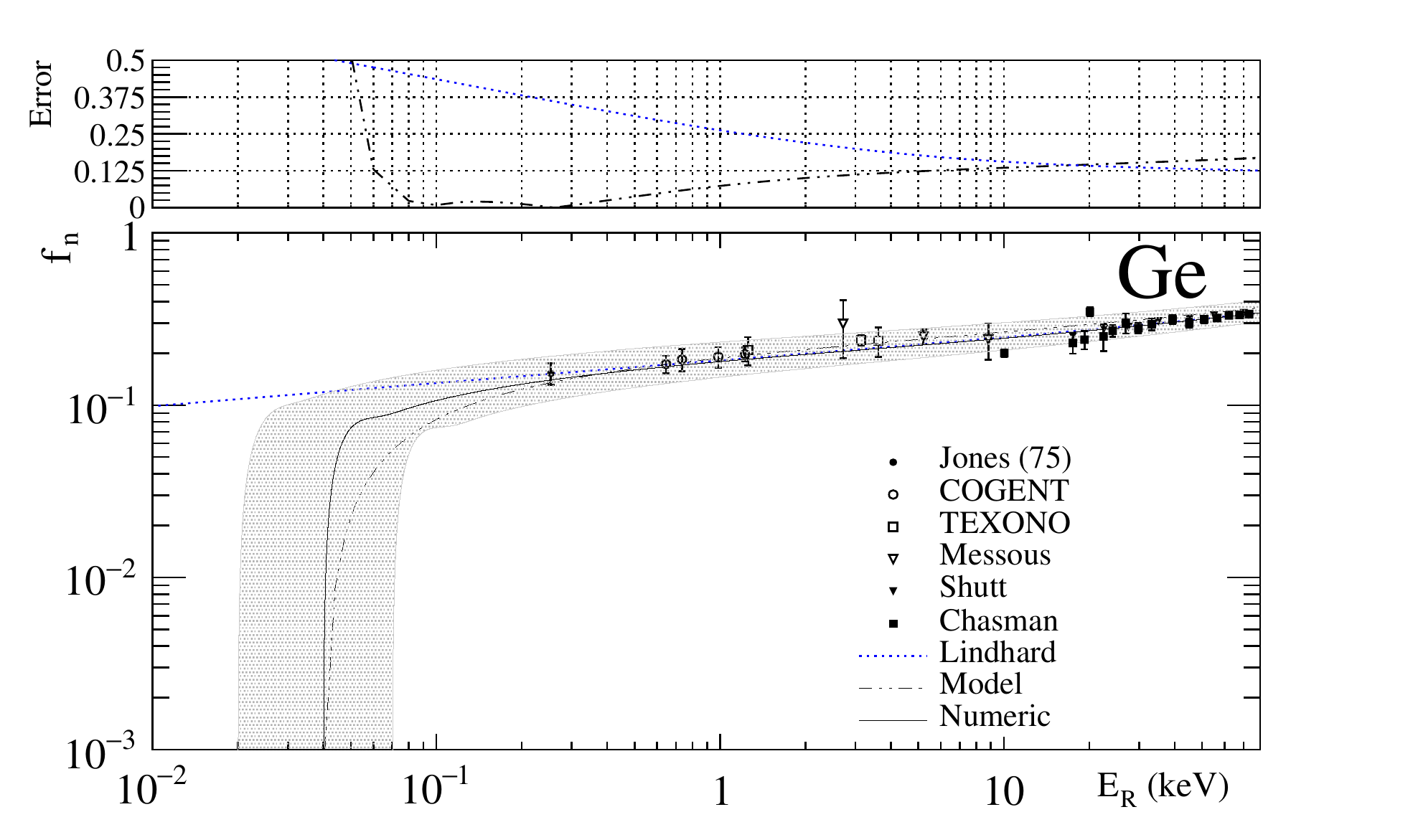}
\caption{ 
(Lower panel) Measurements of the QF in Ge (points with error bars) compared to the Lindhard model (dot-dashed line), the fitted ansatz of Eq.(\ref{eq:corrected_nu_v0}), and the numerical solution with $U=0.02$ keV and $k=0.162$. (Upper panel) Error in the ansatz and the Lindhard original model.
}
\label{fig:fit-Ge}
\end{figure}

The ansatz, Eq.(\ref{eq:corrected_nu_v0}), with $\varepsilon=\varepsilon_R-u$, was fit to the data for each target ion allowing $C_0$, $C_1$, and $u$ to vary freely, with the constraint that the QF displays a cut-off in a positive value of $E_R$. The numerical solution was also fit to the data varying the parameters $k$ and $u$.
The results of the fits are summarized in Table~\ref{tab:fitted-ansatz} for the ansatz, and  Table~\ref{tab:fitted-numeric} for the numerical solution, as well as in Fig.~\ref{fig:fit-Si} for Si and Fig.~\ref{fig:fit-Ge} for Ge. Additionally, we tabulated the numerical solution for the function $f_{n}$ against the recoil energy for Si and Ge in table I and II see supplemental material. The top panel in these figures shows the error calculated using Eq.~(\ref{eq:error-metric}) for the ansatz, and compares it with the error for Lindhard's model tested against his original integral equation, Eq.~(\ref{Eq:SimpIntEq}). By construction the error of the numerical solution is negligible ( $<$ 0.5\%) and is not shown.

The fits of the ansatz and the numerical solution give high values of $\chi^2$ per degree of freedom for Si and Ge, which are indicative of the tension among the different data sets. The uncertainties that we report in Tables \ref{tab:fitted-ansatz} and \ref{tab:fitted-numeric} were estimated so as to approximately cover the variation among the different measurements, and in the case of Xe, to cover the large uncertainties reported. This is shown in the error bands in Figs.~\ref{fig:fit-Si} and \ref{fig:fit-Ge}. 

 For Si data, the ansatz fit (see Table \ref{tab:fitted-ansatz}) gives a value of the binding energy of $U= 0.15 \pm 0.06$ keV, while the fit of the numerical solution (see Table \ref{tab:fitted-numeric}) gives $k=0.161 \substack{+0.029 \\ -0.020} $, and $U=0.15 \substack{+0.10 \\ -0.05}$ ~keV. The fitted value of $k$ is well within the expected values extracted from the older data in the range from 10-100 keV fitted to Lindhard's model. On the other hand, the fitted binding energy is consistent with a picture where the recoiling ion causes, on average, the ionization of one electron from the $2p$ shell, as well as the creation of several $e-h$ pairs and Frenkel pair defects. The cut-off of the QF at $E_r\approx 300$ eV is an artifact of the constant $u$ model arising from the relatively high value of the binding energy, compared to the energy required to produce $e-h$ pairs or lattice defects in Si, which limits the applicability of the model to  $E_r\gtrsim 500$  eV. 

For the Ge data, the ansatz fit gives a value of $U= 0.02\pm 0.01$~keV, and the fitted numerical solution gives $k=0.162 \substack{+0.017 \\ -0.024}$, and $U=0.02\substack{+0.015 \\ -0.010}$~keV. Once more, the fitted value of $k$ agrees well with previous estimates, since the available data can be described reasonably well by Lindhard's original model. Interestingly, since in this case the binding energy is of the same order of magnitude as the energy required to create lattice defects, a naive picture can be considered. The recoiling ion can cause, either  the ionization of one electron from the $3d$ shell, as well as a few $e-h$ pairs, or instead, the creation of one Frenkel-pair and several $e-h$ pairs. The cut-off of the QF from the numerical solution appears at $E_r\approx 40$~eV, which is likely closer to the physical threshold for this target atom. In this case, our constant $u$ model is expected to give a reasonable description all the way down to recoil energies of $E_r\gtrsim 50$~eV, much closer to the physical threshold, which can be safely expected to lie somewhere between a few eV and a few tens of eV.

\begin{figure}[b]
\includegraphics[scale=0.43]{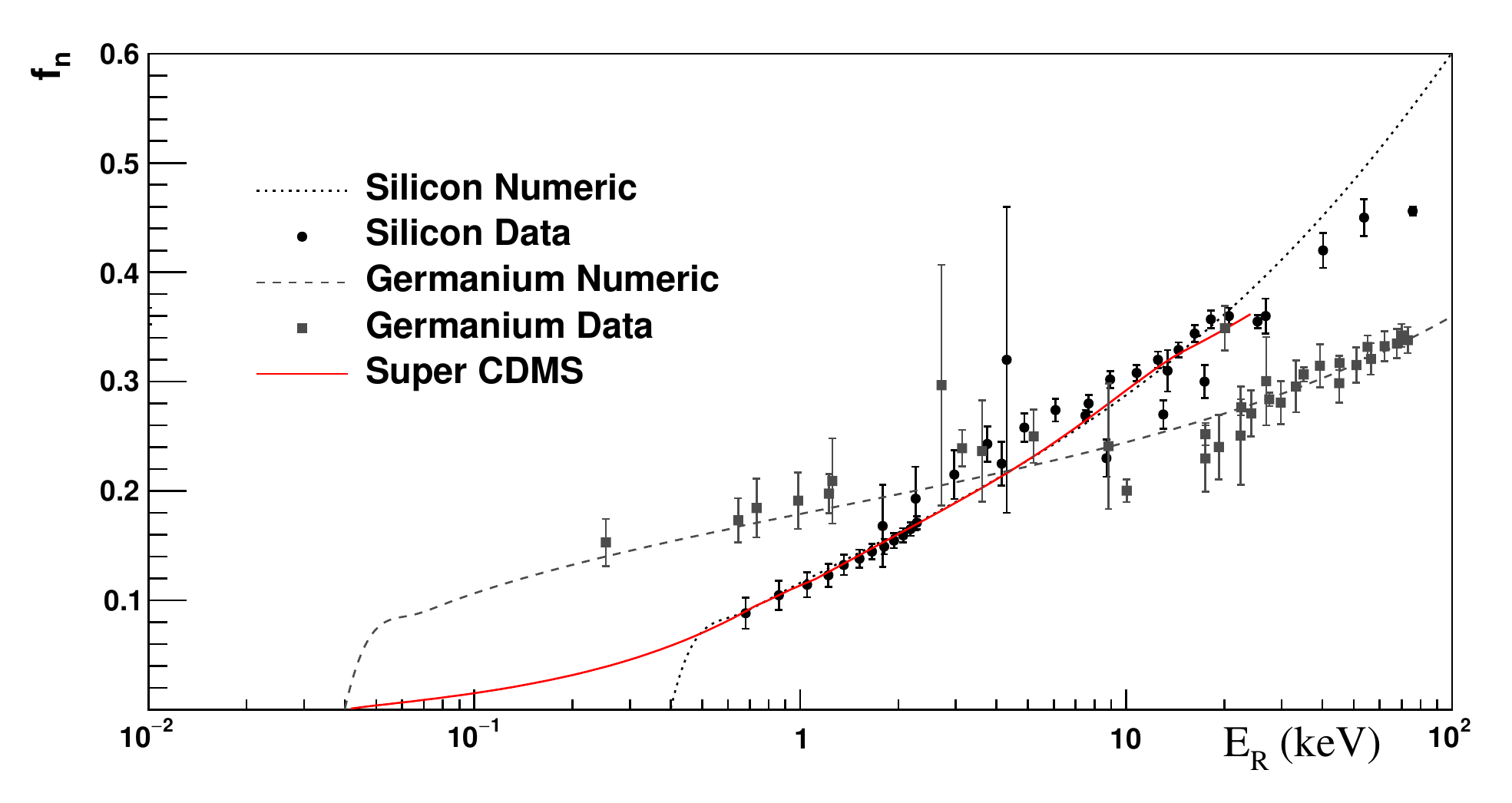}
\caption{ Comparison of the numerical solutions for Si, Ge, with data. The Si curve has been changed from that in Fig.\ref{fig:fit-Si} to fit only the data $<40$~keV ($k=0.169$ and $U=0.2$~keV). Also shown is the phenomenological fit by Super-CDMS \cite{agnese:2017} (red solid line).}
\label{fig:numeric-summary}
\end{figure}
\noindent 
Although the ansatz gives a reasonable description of the data, the numerical solution does so too using only two parameters, and is therefore preferred.
\noindent 
Fig.~\ref{fig:numeric-summary} shows a comparison of the numerical solutions obtained for the three targets considered in this work. In this figure, we have modified the numerical solution for Si to provide a good match to the data below 40~keV, which follows very closely the phenomenological fit reported by the Super-CDMS Collaboration \cite{CDMS}, shown in the solid red line in the figure. The three Si measurements above this energy are likely affected by nuclear charge screening effects, as is suggested by the change in behavior already seen in the Super-CDMS fit.

\section{Conclusions}
\label{sec:conclusions}

\noindent 
We found an appropriate form for the basic integro-differential equation describing the energy given to atomic motion by nuclear recoils in a homogeneous medium, such as pure crystals, when the binding energy is taken into account.
Assuming a constant average binding energy, $u\neq 0$, we give approximate semi-analytical solutions, motivated by the analysis of the integro-differential equation, that are in reasonable agreement with the available experimental measuremens of the QF for nuclear recoils in Si and Ge.
Numerical solutions depending only on the constant binding energy and the electronic stopping power factor $k$ were calculated and found to be also consistent with the data.
As expected, our solutions for the QF display a cut-off at a value equal to twice the binding energy, $2u$. This cut-off is a feature due to the threshold of the cascading process built into the model.

Measurements of QF in Ge detectors are well described by the our model, with $k$ within the expected range ($0.1<k<0.2$). We predict that the QF cut-off in this material is in the range between 20-70 eV  of nuclear recoil energy, corresponding to a binding energy of 10-35 eV. The Frenkel pair dislocation energy in Ge falls well within this interval, and is expected to be an upper limit close to the physical cut-off, believed to be of the order of only a few eV. In a more realistic scenario, where the ion is only required to acquire sufficient motion to generate phonon excitations that can then take an electron from the valence to the conduction band, such a low physical cut-off could be explained.

In the case of Si, the QF measurements are well described by our model with $k$ within the expected limits, only if the binding energy is in the range 100-250~eV. Now, the predicted cut-off is much larger than the Frenkel energy of about 36~eV, and therefore also greater than the physical cut-off. Hence, the model should be valid only for nuclear recoil energies above 500~eV. A more accurate description, considering the variation of the binding energy and stopping power with the recoiling ion energy could be considered.  In addition, effects appearing at higher energies, such as ion charge screening ({\it e.g.}, Bohr stripping \cite{bohrstripping}) could manifest as a change in the value of $k$ at recoil energies of a few tens of keV.

In summary, the model described here, depending only on a constant binding energy and a slope of the velocity-proportional electronic energy loss in the range $0.1<k<0.2$, can explain the behavior of the QF measured to date in pure element targets of Si and Ge. We expect the model to give a reasonable approximation to the physical cut-off in cases where the binding energy is lower than or comparable to the Frenkel-pair energy, as it is the case for Ge.

\section*{\label{sec:Ackno}Acknowledgements}

\noindent 
This research was supported in part by DGAPA-UNAM grant number PAPIIT-IN108917, and Consejo Nacional de Ciencia y Tecnolog\'ia (CONACYT) through grant CB-2014/240666.
The authors wish to thank Guillermo Fernandez Moroni and Juan Estrada for useful discussions.

\begin{appendices}

\section{Semi-hard spheres collision} 
\label{App:semi-hard}

The semi-hard sphere model can be used to calculate the minimum scattering angle, and hence the minimum value of $t$. Taking into account the binding energy $u$, the total energy is $\varepsilon+v$, where $v$ is
\begin{equation}\label{Eq:A1}
 v = \begin{cases}
        0 & \text{for r} \in [R, \infty] \\
       -u & \text{for r} \in [R_{0}, R] \\
       \infty  & \text{for r} \in [0, R_{0}] .\\
     \end{cases}
\end{equation}

\begin{figure}[h!]
\centering
\includegraphics[scale=0.55]{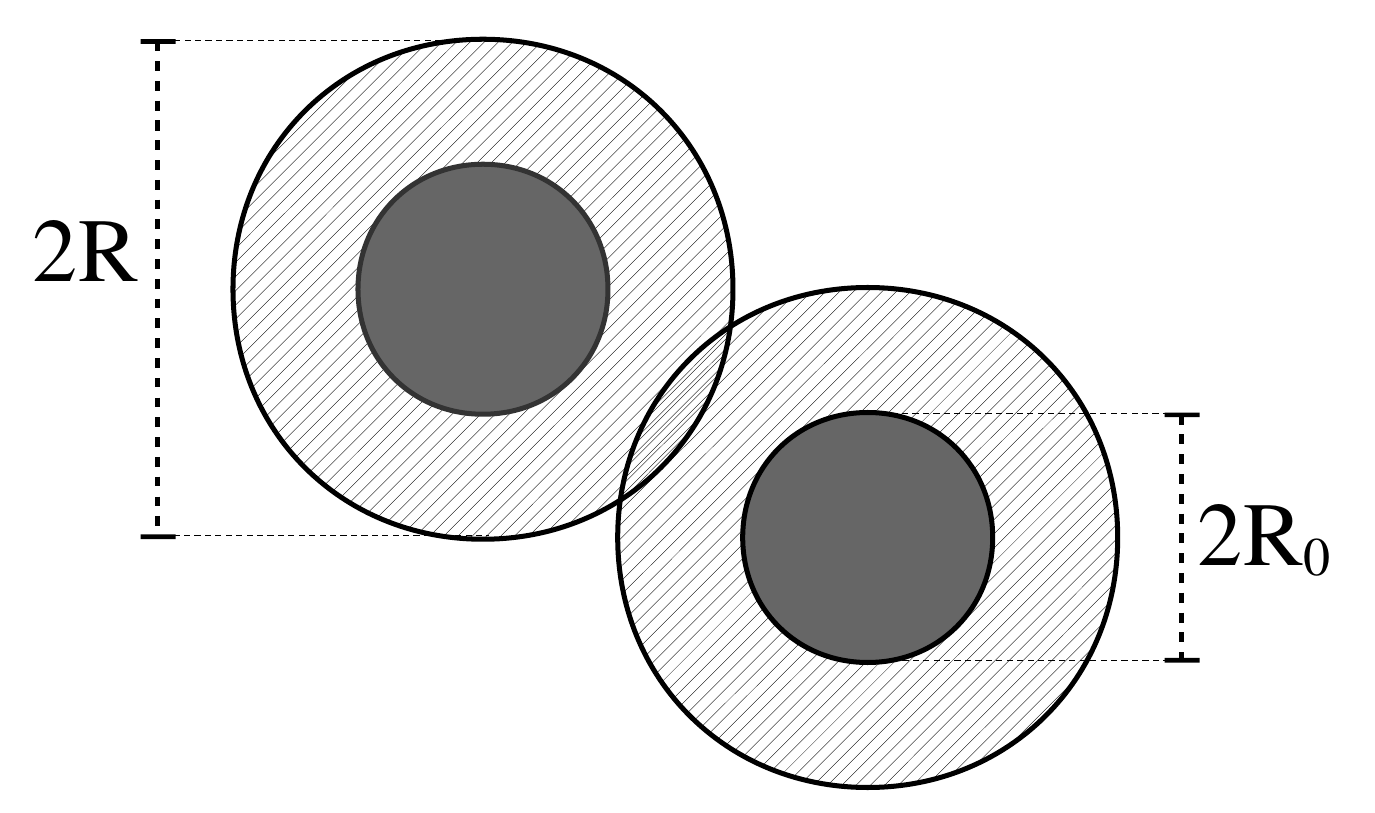}
\caption{ Diagram of a collision between two semi-hard spheres.}
\label{fig:Ap1}
\end{figure}

In order to estimate the minimum scattering angle for this scenario we use as an approximation the classical formula for the scattering angle from a potential 

\begin{equation}\label{Eq:A2}
 \theta_{min}=\pi-2b\int^{\infty}_{r_{min}} \frac{dr}{r^{2}\sqrt{1-(b/r)^{2} -v/(\varepsilon+v)}},
 \end{equation}
where $b$ is the impact parameter (set to $R_{0}$, as shown in Fig.~\ref{fig:Ap1}), 
$r_{min}$ is the turning point of the potential, and $v$ is given in Eq.(\ref{Eq:A1}).

For the potential in Eq.(\ref{Eq:A1}) we can split the integral (\ref{Eq:A2}) in three parts: one from zero to $R_{0}$, another form $R_{0}$ to $R$, and the third from $R$ to $\infty$. The first integral is zero, so the minimum angle is given by 
\begin{eqnarray}\label{Eq:A3}
    \theta_{min}=\pi- \int^{R}_{R_{0}} \frac{2R_{0}\,dr}{r^{2}\sqrt{1-(R_{0}/r)^{2} +u/(\varepsilon-u)}} \nonumber \\
    - \int^{\infty}_{R} \frac{2R_{0}\,dr}{r^{2}\sqrt{1-(R_{0}/r)^{2} }} ,
\end{eqnarray}
 Assuming that $R_{0}\propto a_{0}/Z$, where $a_{0}$ is the Bohr radius and $R\propto 2a_{0}$, for $Z>5$ we have $R\gg R_{0}$, so we can approximate  Eq.(\ref{Eq:A3}) by 
 \begin{eqnarray}\label{Eq:A4}
    \theta_{min} \cong \pi- \int^{\infty}_{R_{0}} \frac{2R_{0}\,dr}{r^{2}\sqrt{1-(R_{0}/r)^{2} +u/(\varepsilon-u)}}.
\end{eqnarray}

Calculating the integral (\ref{Eq:A4}) we arrive at
\begin{equation}\label{Eq:A5}
    \sin^{2}(\theta_{\rm min}/2)=\frac{u}{\varepsilon},
\end{equation}
\noindent
which in terms of the variable $t$ has a minimum at $t_{\rm min}=u\varepsilon$, as is used in Eq.~(\ref{Eq:ModSimIntEq}).
The same result can be derived from the model in Ref.~\cite{Vladimir} (page 131) adapted to the collision of semi-hard spheres.
 


\section{Second order term in the modified simplified integral equation} \label{App:Deriv:ModSimIntEq}

Substitution of Eq.(\ref{Eq:phiExp}) in Eq.(\ref{Eq:BasIntEq}) and integration over the nuclear and electronic cross sections, putting also in effect approximation (D), leads to the appearance of the electronic stopping power
\begin{eqnarray}
    \int d\sigma_{n,e} \; \bar\nu'(E)(\Sigma_i T_{ei}) &=&
    \nu'(E) \int d\sigma_e (\Sigma_i T_{ei}) \nonumber \\
    &\propto & \nu'(\varepsilon)S_e(\varepsilon)
\end{eqnarray}
\noindent in the first derivative term, as in the original formulation by Lindhard. In the second derivative term, we can apply the integral mean value theorem to write
\begin{eqnarray}
    \int d\sigma_{n,e} \; \bar\nu''(E)T_n(\Sigma_i T_{ei}) &=&
    \nu''(E) \, \bar{T}_n \int d\sigma_e (\Sigma_i T_{ei}) \nonumber \\
     \hspace{3cm}
     &\propto& \nu''(\varepsilon) \, \bar{t}_n  S_e(\varepsilon),
\end{eqnarray}
\noindent where $\bar{t}_n = c\bar{T}_n$ is a suitable average value of the energy transfer $t_n=\varepsilon\sin^2(\theta/2)$, which we will approximate by $\bar{t}_n\approx \langle t_n \rangle = \tfrac{1}{2}\varepsilon$, leading to the final form of our proposed modified simplified integral equation Eq.~(\ref{Eq:ModSimIntEq}).

\end{appendices}

\newpage
\bibliography{QFpaper}
\end{document}